# Strong enhancement of the spin Hall effect by spin fluctuations near the Curie point of Fe$_x$Pt$_{1-x}$ alloys

Yongxi Ou[1*], D.C. Ralph[1,2] and R.A. Buhrman[1*]

[1]Cornell University, Ithaca, New York 14853, USA,

[2]Kavli Institute at Cornell, Ithaca, New York 14853, USA

*email: yo84@cornell.edu; rab8@cornell.edu.

## Abstract

Robust spin Hall effects (SHE) have recently been observed in non-magnetic heavy metal systems with strong spin-orbit interactions. These SHE are either attributed to an intrinsic band-structure effect or to extrinsic spin-dependent scattering from impurities, namely side-jump or skew scattering. Here we report on an extraordinarily strong spin Hall effect, attributable to spin fluctuations, in ferromagnetic Fe$_x$Pt$_{1-x}$ alloys near their Curie point, tunable with *x*. This results in a damping-like spin-orbit torque being exerted on an adjacent ferromagnetic layer that is strongly temperature dependent in this transition region, with a peak value that indicates a lower bound $0.34 \pm 0.02$ for the peak spin Hall ratio within the FePt. We also observe a pronounced peak in the effective spin-mixing conductance of the FM/FePt interface, and determine the spin diffusion length in these Fe$_x$Pt$_{1-x}$ alloys. These results establish new opportunities for fundamental studies of spin dynamics and transport in ferromagnetic systems with strong spin fluctuations, and a new pathway for efficiently generating strong spin currents for applications.

The manipulation of the magnetization in ferromagnetic (FM) nanostructures with pure spin current densities $J_s$ has become a primary tool in spintronics since the demonstrations of nanomagnet switching driven by large spin orbit torques (SOT) originating from the spin Hall effect (SHE) [1–4] in an adjacent heavy metal carrying a longitudinal electrical current density $J_e$. Most SOT efforts so far have focused on the utilization of large *intrinsic* spin Hall ratios, $\theta_{SH} \equiv (2e/\hbar) J_s / J_e$, for certain heavy metals that are compatible with the requirements of a successful spintronics technology [5–9]. An alternative approach to enhance $\theta_{SH}$ is to introduce dopants into a metallic system whereby strong spin-orbit interactions can strenghten the spin Hall effect, either by enhanced *extrinsic* spin-dependent skew or side-jump scatteringor by the *intrinsic* effect through a beneficial modification of the electronic band structure at the Fermi level [10]. Typically the dopant has been a heavy element without a strong magnetic moment, e.g. Ir, Bi, Au [11–14], and the resulting enhancement, typically attributed to skew-scattering, has been measured by the inverse spin Hall effect or by a non-local spin accumulation technique. Recently Wei et al. [15] have reported a moderate, but notable, temperature-dependent enhancement of the inverse spin Hall effect in dilute NiPd alloys, attributed to spin-fluctuation-enhanced skew scattering by the Ni ions in the vicinity of the ferromagnetic transition. We note that NiPd is one of a class of ferromagnetic alloys that have long been know to exhibit giant magnetic moments per ferromagnetic solute, particularly in the dilute limit [16]. Reference [15] provides strong motivation for examining the direct SHE in other ferromagnetic alloys in which there can be a stronger spin-orbit interaction between the conduction electrons and the ferromagnetic component. Here we demonstrate that for Fe-doped Pt alloys, $Fe_xPt_{1-x}$, the effective spin Hall angle as measured directly from the damping-like torque exerted on an adjacent ferromagnetic layer is increased by more than a factor of 3 in the vicinity of its Curie temperature $T_c$ in comparison to the already-substantial value it has well above $T_c$, and at its peak has a value at least comparable to that of beta-W [5], and with a much lower electrical resistivity.

$Fe_xPt_{1-x}$ alloys are well known for their unusually robust magnetic anisotropy properties arising from the strong conduction electron spin-orbit interaction with the Fe orbital moment [17,18], and also for the dependence of the magnetic state on the chemical order. For example, well-ordered $Fe_{0.25}Pt_{0.75}$ exhibits antiferromagnetism, while chemically disordered $Fe_{0.25}Pt_{0.75}$ is ferromagnetic [19,20]. Ferromagnetic $Fe_xPt_{1-x}$ films also exhibit quite large anomalous Hall effects (AHE) [21–23], which suggests that $Fe_xPt_{1-x}$ can be a promising material for the generation of spin currents by the extrinsic SHE.

To investigate this possibility we prepared multilayers containing two different sets of $Fe_xPt_{1-x}$ thin films made by co-deposition at room temperature via dc magnetron sputtering; in one case the nominal composition was $Fe_{0.15}Pt_{0.85}$ and in the other $Fe_{0.25}Pt_{0.75}$. Multilayer stacks consisting of substrate/Ta/IrMn/$Fe_xPt_{1-x}$/MgO/Ta were used for thin film characterization and substrate/Ta/IrMn/$Fe_xPt_{1-x}$/Hf/FeCoB/Hf/MgO/Ta stacks were used for the SOT measurements. These samples were prepared by direct current (dc) sputtering (with rf magnetron sputtering for the MgO layer) in a deposition chamber with a base pressure $<5\times10^{-8}$ Torr. The dc sputtering condition was 2mTorr Ar pressure. The $Fe_xPt_{1-x}$ alloy was grown by co-sputtering from two pure sputtering targets (i.e. Fe and Pt targets). All samples in this work had a Ta(1nm) seed layer to provide a smooth base layer and a Ta(1nm) top layer to provide an oxidized protection layer for the stack. All samples were annealed in an in-plane magnetic field (2000 Oe) in a vacuum furnace at 300C for 1 h to enhance the PMA. For measurements of the AHE and SOT, Hall bar devices with lateral dimensions of $5\times60\,\mu m^2$ were patterned via photolithography and ion milling (see the sample schematic in Fig. 1d).

We performed x-ray diffraction (XRD) measurements on two multilayer samples with the layer structures: $IrMn_3(10)/Fe_{0.15}Pt_{0.85}(10)/MgO$ and $IrMn_3(10)/Fe_{0.25}Pt_{0.75}(10)/MgO$ (the number in parentheses is the thickness in nanometers). The $IrMn_3$ layer was included to provide antiferromagnetic pinning of the $Fe_xPt_{1-x}$ layers when cooled to well below their Curie points for research that will be presented elsewhere; the $Fe_xPt_{1-x}$ layers are thick enough that in the experiments to be considered below the $IrMn_3$ does not contribute any significant SOT on the free magnetic layer. We show in Fig. 1a the (111) XRD peaks for the $Fe_{0.15}Pt_{0.85}$ and the $Fe_{0.25}Pt_{0.75}$ samples, and also for separate 10 nm Pt, $Fe_{0.50}Pt_{0.50}$ and IrMn films for comparison. The (111) peak is reasonably narrow, shifting to higher $2\theta$ angle as the Fe content increases, indicating a decrease in the unit cell size with increased Fe content. As expected from the use of room temperature deposition there was no evidence of a (110) peak in the XRD of these samples that would indicate significant chemical order [24] (The small peak at $2\theta \approx 41^o$ in Fig.1a is due to the $IrMn_3$ base layer). Finally as expected for a disordered metal the resistivity of the films was only weakly temperature dependent, decreasing by less than 10% from room temperature to 160 K, indicating the dominance of impurity scattering. The resistivity of the films did vary with Fe content, from $\rho_{Pt}(10) \approx 15 \mu\Omega\cdot cm$ to $\rho_{Fe_{0.15}Pt_{0.85}}(10) \approx 55 \mu\Omega\cdot cm$ to $\rho_{Fe_{0.25}Pt_{0.75}}(10) \approx 75 \mu\Omega\cdot cm$, indicating an increased electron scattering rate with increased Fe content.

To further confirm the chemical disorder and the ferromagnetic character of these alloys we made temperature-dependent vibrating sample magnetometry (VSM) measurements of the samples (Fig. 1b). Both $Fe_{0.15}Pt_{0.85}$ and $Fe_{0.25}Pt_{0.75}$ were found to be ferromagnetic at sufficiently low temperature, with fits of the spontaneous magnetization $M_s(T)$ to the empirical function $M_s(T) = M_s(0) \cdot (1 - (T/T_c)^\alpha)^\beta$ [25] yielding a Curie temperature of $T_c \approx 174\,K$ for the Fe$_{0.15}$Pt$_{0.85}$ sample and $T_c \approx 288\,K$ for the Fe$_{0.25}$Pt$_{0.75}$ sample.

In Fig. 1c we show the temperature dependence of the "anomalous Hall angle" = $\rho_{xy}/\rho_{xx}$ of the Fe$_{0.15}$Pt$_{0.85}$(10) and Fe$_{0.25}$Pt$_{0.75}$(10) samples as measured in a magnetic field $H_z$ = 2 kOe applied perpendicular to the plane of the film. (Here $\rho_{xx}$ is the resistivity in the direction of current flow and $\rho_{xy}$ is the transverse Hall resistivity). As can be seen in the Fe$_{0.15}$Pt$_{0.85}$(10) sample, there is a significant AHE at high temperature that increases gradually as the temperature is decreased toward 200 K and then increases more rapidly as the magnetization in the film develops as $T$ decreases below $T_c$, qualitatively as would be expected for the case of strong skew scattering from the Fe ions. In the inset of Fig. 1c, there is a similar temperature dependence trend for the Fe$_{0.25}$Pt$_{0.75}$(10) sample below its Curie temperature. In the AHE what is detected is the charge flow in the direction perpendicular to the plane defined by the bias current direction **y** and the direction of the internal magnetic field **z**. This transverse charge flow is accompanied by a diffusive spin current arising from the spin-dependent scattering. The resulting $V_{AHE}$, or equivalently $\rho_{xy}$, scales, for the extrinsic case, with the rate of skew scattering, but it also depends on the strength of the internal magnetization of the material and its spin dependent charge transport properties. This makes it challenging to quantify the underlying spin flow based only on AHE measurements.

To achieve better quantitative measurements of the spin currents produced by an electrical current in the FePt alloys we employed the harmonic response SOT technique [26,27] whereby we measured the magnetic deflection of an adjacent, perpendicularly magnetized ferromagnetic thin film that occurs as the result of the spin torque arising from the absorption of the transverse polarized component of the spin current emanating from the spin source, the FePt alloys in this case. Such measurements of SOT effective fields usually only set a lower bound on $q_{SH}$ due to the expected less than perfect spin transparency of the interface between the spin source and spin sink [28,29].

For the harmonic response SOT measurements we fabricated two sets of FePt-based multilayer samples IrMn$_3$(10)/Fe$_{0.15}$Pt$_{0.85}$(10)/Hf(1)/FeCoB(1)/Hf(0.35)/MgO (A), and IrMn$_3$(10)/Fe$_{0.25}$Pt$_{0.75}$(10)/Hf(0.8)/FeCoB(1)/Hf(0.35)/MgO (B), where FeCoB represents Fe$_{60}$Co$_{20}$B$_{20}$. The amorphous Hf insertion layer (1nm and 0.8 nm for (A) and (B) respectively) between the FePt and the FeCoB was employed to counter the detrimental effect of the fcc crystal structure of the FePt on obtaining perpendicular magnetic anisotropy (PMA) in the thin FeCoB layer, while the very thin (0.35nm) Hf insertion layer between the FeCoB and the MgO enhanced the interfacial magnetic anisotropy energy density, strengthening the PMA [30].

In Fig. 2a, we show the response of the anomalous Hall resistance of one of the Fe$_{0.15}$Pt$_{0.85}$ heterostructure Hall bars (sample A) to an applied out-of-plane field $H_z$ at different temperatures between 300 K and 140 K. The sharp field-induced switching events, with an increasing coercivity upon decreasing temperature, are from the PMA FeCoB layer. When the temperature is lower than the Curie temperature of Fe$_{0.15}$Pt$_{0.85}$ there is also a quasi-linear background evident for $H_z$ greater than the coercive fields that is much larger than can be expected from the ordinary Hall effect, and instead is due to the AHE of the in-plane magnetized FePt layer. The AHE resistance for sample (B) is similar [31].

We determined the damping-like (DL) and field-like (FL) effective fields ($\Delta H_{DL}$ and $\Delta H_{FL}$ respectively) arising from the SOT by measuring the first and second harmonic transverse

Hall signals $V_\omega$ and $V_{2\omega}$ [27], from which we obtain $\Delta H_{L(T)} = -2(\partial V_{2W}/\partial H_{L(T)})/(\partial^2 V_W/\partial H_{L(T)}^2)$ and hence $\Delta H_{DL} = (\Delta H_L + 2\delta \cdot \Delta H_T)/(1-4\delta^2)$, and $\Delta H_{FL} = (\Delta H_T + 2\delta \cdot \Delta H_L)/(1-4\delta^2)$. Here $H_{L(T)}$ is the external bias field applied parallel to (transverse to) the current direction and $\delta$ is the ratio of the planar Hall resistance to the anomalous Hall resistance [26,27].

In Fig. 2b we show the temperature dependences of the DL effective fields for both sample (A) and sample (B), plotted as a function of $T/T_c$, with $T_c$ determined from the fits to the magnetization of the samples (Fig. 1b) (See [31] for discussion of the FL SOT behavior of these samples). For sample (A) ($Fe_{0.15}Pt_{0.85}$) the measurement is from room temperature 293 K to $T$ = 160 K, and for sample (B) ($Fe_{0.25}Pt_{0.75}$) from 330 to 275 K. For sample (A) for which we have measurements starting around 100 degrees above $T_c$, we see that the DL effective field per current density $\Delta H_{DL}/\Delta J_e$ is more or less constant until $T/T_c \approx 1.44$ (250 K), below which it increases, at first gradually, then very rapidly reaching a peak near $T_c$ (172K) more than 3 times its 293 K value. This behavior is dramatically different from that of the DL torque found with conventional heavy metal systems [32,33]. Below this peak $\Delta H_{DL}/\Delta J_e$ then drops off even more quickly until below $T/T_c = 0.92$ (160 K) the observable development of spin torque from the PMA FeCoB on the emerging strong ferromagnetism of the in-plane polarized $Fe_{0.15}Pt_{0.85}$ makes further quantitative harmonic response measurements untenable (see [31] for more information).

As also shown in Fig. 2b the behavior of sample (B) over the same scaled temperature range above and below $T_c$, is quite similar, with the peak value of $\Delta H_{DL}/\Delta J_e$ being less that 20% different than that of sample (A), and even less if we take into account the spin attenuation effect of the different Hf spacer thickness (1 nm for sample A and 0.8 nm for sample B) where Hf has a spin diffusion length of approximately 1 nm [34]. This close similarity in the values of the peak anti-damping spin torque is observed despite the 35% difference in resistivity, and 67% difference in Fe concentration. This is consistent with skew scattering being the dominant spin Hall effect in these materials, at least in the ferromagnetic transition region, but more study will be needed to confirm that attribution.

Some years ago Kondo [35] developed a theory for the scattering of conduction electrons by localized orbital moments to explain an anomaly in the magnetoresistance and AHE of ferromagnetic Ni and Fe near their Curie points [36], with Kondo attributing the anomaly to increasingly stronger spin fluctuations as $T \to T_c$ from below. Recently Gu *et al.* [37] extended this theory to explain results by Wei *et al.* [15] on inverse spin Hall effect (ISHE) measurements of NiPd alloys near their $T_c$, including the effect of correlations between neighboring localized moments. We surmise that spin fluctuations are also the origin of the strong peak in the SOT torque (spin current) that we observe with the FePt alloys, although our results, in addition to being a direct measure of the $J_s$ generated by the spin Hall effect, differ from the earlier work by the strength of the effect, which we attribute to the exceptionally strong spin-orbit interaction in the Fe-Pt system. Our results are also distinctive in that the peak in the SOT effective field (emitted spin current) is followed by a sharp decline with decreasing temperature that we attribute to the effect of the internal exchange field in the FePt that develops as $T$ is lowered below $T_c$, which, once well established, acts to quickly dephase a spin current that is polarized in a direction not collinear with the internal magnetization [23].

If spin fluctuations are indeed the origin of the enhanced SHE in the FePt alloys near $T_c$, then it is predicted [38] that there should also be an enhancement of the effective interfacial spin-mixing conductance $g_{eff}^{\uparrow\downarrow}$ between the FeCoB and FePt alloy in the vicinity of the latter's Curie point, as has been recently observed with *antiferromagnetic* spin sinks near their Neel point by inverse spin Hall measurements [39] and spin pumping [40]. In the latter case the interfacial enhancement of damping $\Delta\alpha \equiv \alpha(t_{FM}) - \alpha_0$, where $\alpha_0$ is the Gilbert damping parameter for the bulk FM material, can be related to $g_{eff}^{\uparrow\downarrow}$ by $g_{eff}^{\uparrow\downarrow} = 4\pi M_s t_{FeCoB} \Delta\alpha / (\gamma\hbar)$ [41]. We do indeed observe a pronounced peak in magnetic damping of the FeCoB layer in our samples as they are cooled through the $T_c$ of the FePt. In Fig. 3a we show the effective spin mixing conductance, $g_{eff}^{\uparrow\downarrow}$ as determined by resonant linewidth measurements made by flip-chip field-modulated FMR, on a Fe$_{0.25}$Pt$_{0.75}$(10)/Hf(0.25)/FeCoB(7.3)/MgO sample (See [31] for details of measurements). As can be seen $g_{eff}^{\uparrow\downarrow}$ increases rapidly as $T$ moves below $T_c$, and then drops abruptly by more than a factor of three to a value ($\approx 30 \, \text{nm}^{-2}$) much closer to that expected for a typical FM/Pt interface [29]. The temperature-dependent behavior observed here is distinctly different from the temperature-insensitive $g_{eff}^{\uparrow\downarrow}$ in Pt/ferromagnet bilayer systems [42]. Note that the peak in $g_{eff}^{\uparrow\downarrow}$ does not occur simultaneously with the peak in $\Delta H_{DL}/\Delta J_e$ which indicates that the latter's peak is not just due to an enhanced $g_{eff}^{\uparrow\downarrow}$.

To better quantify the peak strength of SHE in the FePt alloys and to account for the spin current attenuation in the Hf spacer layer, we prepared another Fe$_{0.25}$Pt$_{0.75}$ heterostructure, IrMn$_3$(10)/Fe$_{0.25}$Pt$_{0.75}$(10)/Hf(0.5)/FeCoB(0.9)/Hf(0.35)/MgO, i.e. with a thinner (0.5nm) Hf spacer layer. In Table I we compare the peak $\Delta H_{DL}/\Delta J_e$ value that we obtained with a Hall bar measurement of this sample to that previously measured [33] for a Pt(4)/Hf(0.5)/FeCoB(1)/MgO sample at 293 K (room temperature). Assuming the same spin current attenuation in both cases from the 0.5 nm Hf/FeCoB interface this indicates that the peak spin Hall effect in the $Fe_{0.25}Pt_{0.75}$ is approximately 5.5x larger than in Pt(4). More quantitatively, with the torque-field relation $\xi_{DL} = \frac{2e}{\hbar} M_s t_{FeCoB} \frac{\Delta H_{DL}}{\Delta J_e}$ [29], we can calculate the DL spin torque efficiency $= \xi_{DL} = 0.34 \pm 0.02$ for the sample $Fe_{0.25}Pt_{0.75}(10)/Hf(0.5)/FeCoB(1)$. Considering the attenuation of the spin current as it passes through the 0.5 nm of Hf, and the likely less than ideal spin transparency of the Hf/FeCoB interface, this only sets the *lower bound* for the peak spin Hall ratio of the $Fe_{0.25}Pt_{0.75}$ material as $\geq 0.34$.

Another key parameter for understanding and optimizing the effectiveness of SOTs arising from the SHE is the spin diffusion length $\lambda_s$ within the material. We obtained a measure of $\lambda_s$ by producing a series of PMA samples without the IrMn layer $Fe_{0.25}Pt_{0.75}(t_{FePt})/Hf(0.8)/FeCoB(1)/Hf(0.35)/MgO/Ta(1)$ where the thickness $t_{FePt}$ of the FePt alloy was varied from 2 to 10 nm. The measured damping-like effective fields for these samples are plotted in Fig. 3b as a function of $t_{FePt}$ for two different temperatures 293 and 330 K, *i.e.* in the near vicinity of $T_c$ and somewhat above it. The solid lines are a fit of the function $\Delta H_{DL}(t_{FePt})/\Delta J_e = (\Delta H_{DL}(\infty)/\Delta J)(1 - \text{sech}(t_{FePt}/\lambda_s))$ [43] to the data. The results at the two temperatures are quite similar, with $\lambda_s \approx 1.5$ nm.

We further confirm the strength of the SHE in the FePt alloys with current-induced switching measurements. In Fig. 4 we show the switching behavior of sample (B) as measured at $293 K$, in close vicinity to $T_c$. The direction of current-induced switching is reversed upon changing the sign of a small in-plane applied magnetic field, in a way that is characteristic of antidamping torque SHE switching [44]. The switching current density for this case was $\approx 6\times 10^6$ A/cm$^2$.

In summary, we have studied the spin-orbit torques resulting from the SHE in chemically disordered FePt alloys, above and through their ferromagnetic transition points $T_c$. The SOTs exerted by these materials on an adjacent FeCoB thin film exhibit a striking temperature dependence in which the DL SOT displays a strong maximum in the vicinity of $T_c$. We attribute this pronounced SOT behavior to spin fluctuation enhancement of the spin Hall effect arising from the strong spin-orbit interaction between the conduction electrons and the localized Fe moments. There is also a strong *T*-dependent enhancement of the effective spin-mixing conductance of the FeCoB/FePt interface that we similarly attribute to spin fluctuations in the FePt ferromagnetic transition region. The peak strength of the DL SOT indicates an exceptionally large spin Hall angle, > 0.34 near the Curie point of the FePt alloys. We also realized current-induced magnetization switching by the DL SOT in close vicinity to the Curie point of these ferromagnetic FePt alloys and measured the spin diffusion length to be quite similar, ≈ 1.5 nm, both above and in close vicinity to $T_c$. This fluctuation enhanced spin Hall effect, which is tunable through the composition of the FePt alloy, provides new opportunities for the study of spin-dependent scattering and transport in systems with very strong spin-orbit interactions, and for applications where a very strong spin current from a relatively low resistivity material can be particularly beneficial.

**Acknowledgements**

Y.O. thanks Shengjie Shi for the assistance in low temperature flip-chip FMR measurements. This research was supported by ONR (N000014-15-1-2449) and by NSF/MRSEC (DMR-1120296) through the Cornell Center for Materials Research (CCMR), and by NSF through use of the Cornell NanoScale Facility, an NNCI member (ECCS-1542081).

**Competing Financial Interests**

The authors hold patents and have patent applications filed on their behalf regarding some aspects of the spin-orbit torque research discussed in this report.

**109**, 096602 (2012).

**Figure 1**

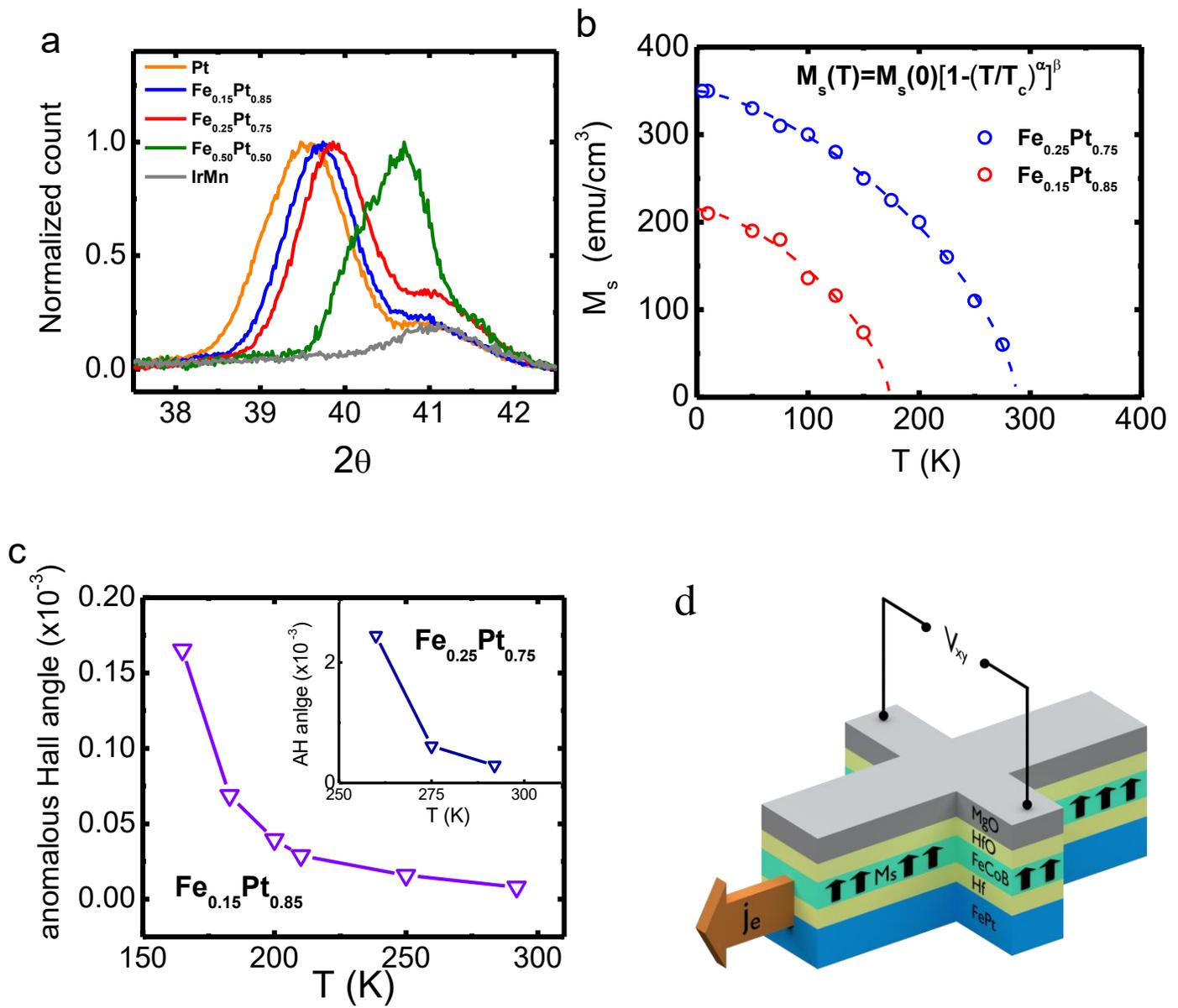

Figure 1. (a) XRD measurements on the samples: IrMn$_3$(10)/F$_{0.50}$Pt$_{0.50}$(10)/MgO, IrMn$_3$(10)/Fe$_{0.25}$Pt$_{0.75}$(10)/MgO and IrMn$_3$(10)/Fe$_{0.15}$Pt$_{0.85}$(10)/MgO, and two control samples: IrMn$_3$(10)/Pt(1)/MgO and IrMn$_3$(10)/Pt(8)/MgO. (b) Temperature dependent VSM measurements on the samples IrMn$_3$(10)/Fe$_{0.25}$Pt$_{0.75}$(10)/MgO and IrMn$_3$(10)/Fe$_{0.15}$Pt$_{0.85}$(10)/MgO. The dashed lines are fits to the empirical equation $M_s(T) = M_s(0) \cdot (1 - (T/T_c)^\alpha)^\beta$. (c) Temperature dependence of the anomalous Hall angle of the samples IrMn$_3$(10)/Fe$_{0.15}$Pt$_{0.85}$(10)/MgO (main) and IrMn$_3$(10)/Fe$_{0.25}$Pt$_{0.75}$(10)/MgO (inset). (d) Schematic of the Hall bar device.

**Figure 2**

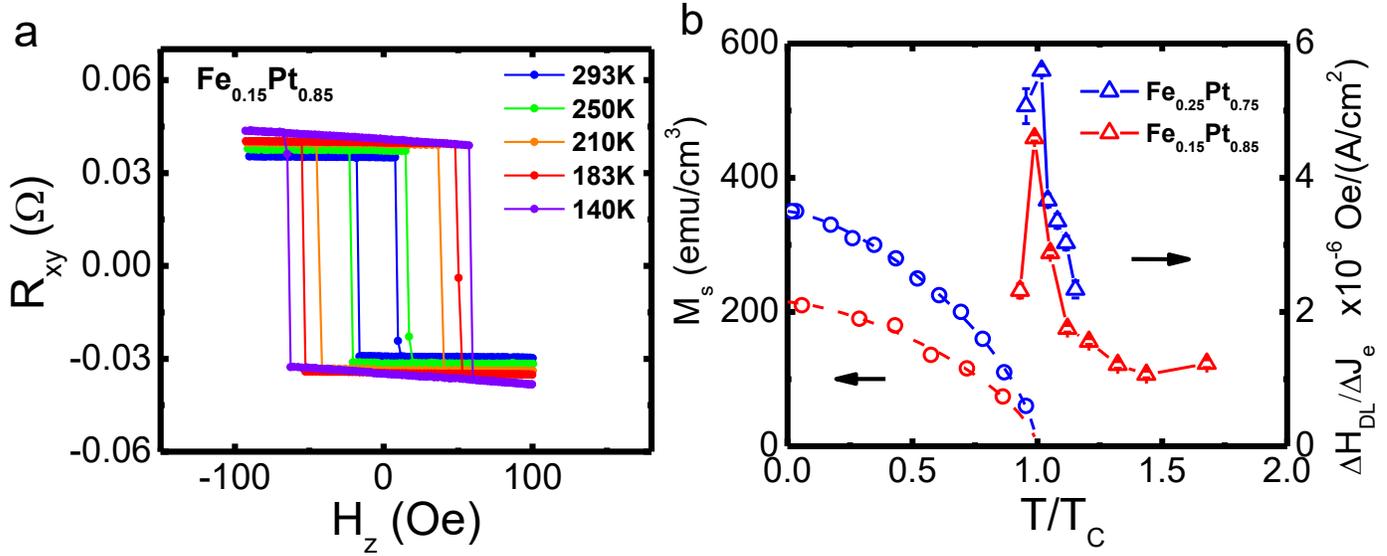

Figure 2. (a) Temperature-dependent AHE resistance of sample (A) Fe$_{0.15}$Pt$_{0.85}$. (b) Dampinglike effective fields of sample (A) Fe$_{0.15}$Pt$_{0.85}$ and (B) Fe$_{0.25}$Pt$_{0.75}$ as a function of normalized temperature $T/T_c$. Their temperature dependent magnetizations are also plotted here for comparison.

**Figure 3**

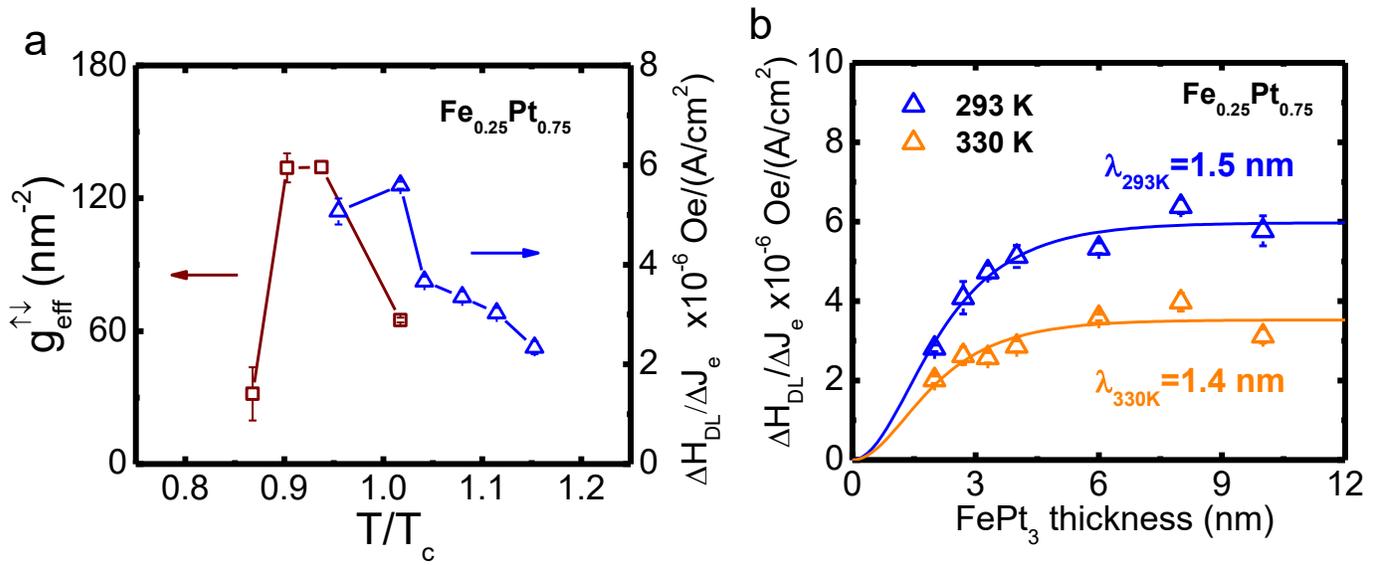

Figure 3. (a) The effective spin mixing conductance of an in-plane magnetized $Fe_{0.25}Pt_{0.75}(10)/Hf(0.25)/FeCoB(7.3)$ sample as determined by a flip-chip FMR measurement of the damping parameter for the FeCoB resonance. The temperature dependence of the DL effective field of sample (B) is also plotted here for comparison. (b) Spin diffusion length measurement of the samples $Fe_{0.25}Pt_{0.75}(t)/Hf(0.8)/FeCoB(1)$.

**Figure 4**

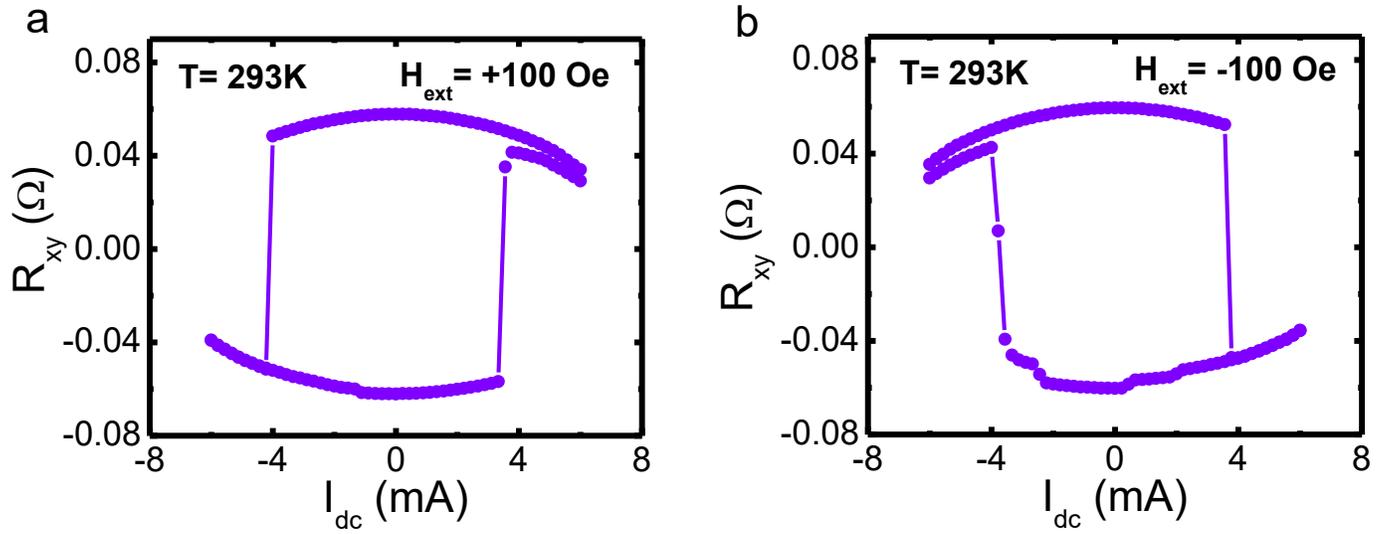

Figure 4. Current-induced magnetization switching of sample (B) IrMn$_3$(10)/Fe$_{0.25}$Pt$_{0.75}$ (10)/Hf(0.8)/FeCoB(1)/Hf(0.35)/MgO at room temperature under an external magnetic field along the current direction.

**Table I**

| Sample | $Fe_{0.25}Pt_{0.75}(10)/Hf(0.8)/FeCoB(1)$ | $Fe_{0.25}Pt_{0.75}(10)/Hf(0.5)/FeCoB(1)$ | $Pt(4)/Hf(0.5)/FeCoB(1)$ |
|---|---|---|---|
| DL effective field $\times 10^{-6}$ Oe/(A/cm$^2$) | 5.6 | 12.2 | 2.3 |
| Reference | This work | This work | Ou et al. [33] |

Table I: Comparison of the dampinglike effective fields as measured at 293 K for two $Fe_{0.25}Pt_{0.75}$/Hf/FeCoB samples and a Pt/Hf/FeCoB sample.

**Strong enhancement of the spin Hall effect by spin fluctuations near the Curie point of $Fe_xPt_{1-x}$ alloys-Supplementary Materials**

Yongxi Ou[1*], D.C. Ralph[1,2] and R.A. Buhrman[1*]

[1]Cornell University, Ithaca, New York 14853, USA,

[2]Kavli Institute at Cornell, Ithaca, New York 14853, USA

*email: yo84@cornell.edu; rab8@cornell.edu.

**Table of contents:**

## S1. Temperature-dependent anomalous Hall measurements of the PMA sample $Fe_{0.25}Pt_{0.75}$ (10)/Hf(0.8)/FeCoB(1)/Hf(0.35)/MgO

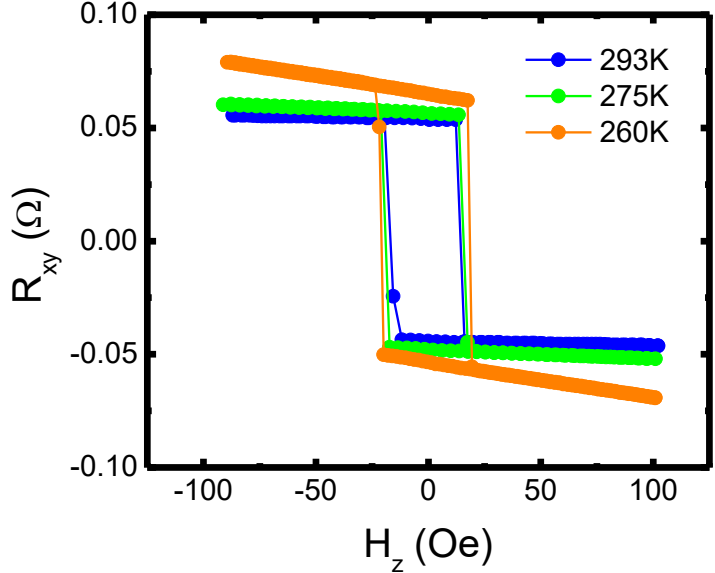

**S1.** Anomalous Hall resistance of $IrMn_3(10)/ Fe_{0.25}Pt_{0.75}$ (10)/Hf(0.8)/FeCoB(1)/Hf(0.35)/MgO at three different temperatures near and below the Curie temperature $T_c = 288$ K of the $Fe_{0.25}Pt_{0.75}$.

Figure S1 shows anomalous Hall (AH) resistance measurements of a Hall bar of the sample (B) heterostructure $IrMn_3(10)/Fe_{0.25}Pt_{0.75}(10)/Hf(0.8)/FeCoB(1)/Hf(0.35)/MgO$ as a function of an external magnetic field applied perpendicular to the plane of the sample, as measured from room temperature (293K) to 260K. Just as for the data from sample (A) discussed in the main text, these anomalous Hall signals have contributions from both the perpendicularly-magnetized FeCoB(1nm) layer and the in-plane magnetized $Fe_{0.25}Pt_{0.75}$(10nm) layer. For the PMA FeCoB layer, the coercive field increases as the temperature $T$ is reduced but there is only a minimal increase in its contribution to the AH resistance. The $Fe_{0.25}Pt_{0.75}$ layer contributes a background that is linear in applied field, with a slope that increases in magnitude as $T$ is decreased below $T_c$ due to the increasing magnetization of the $Fe_{0.25}Pt_{0.75}$ layer.

## S2. Spin-orbit-torque effective fields arising from the FePt alloys as determined by the harmonic response technique

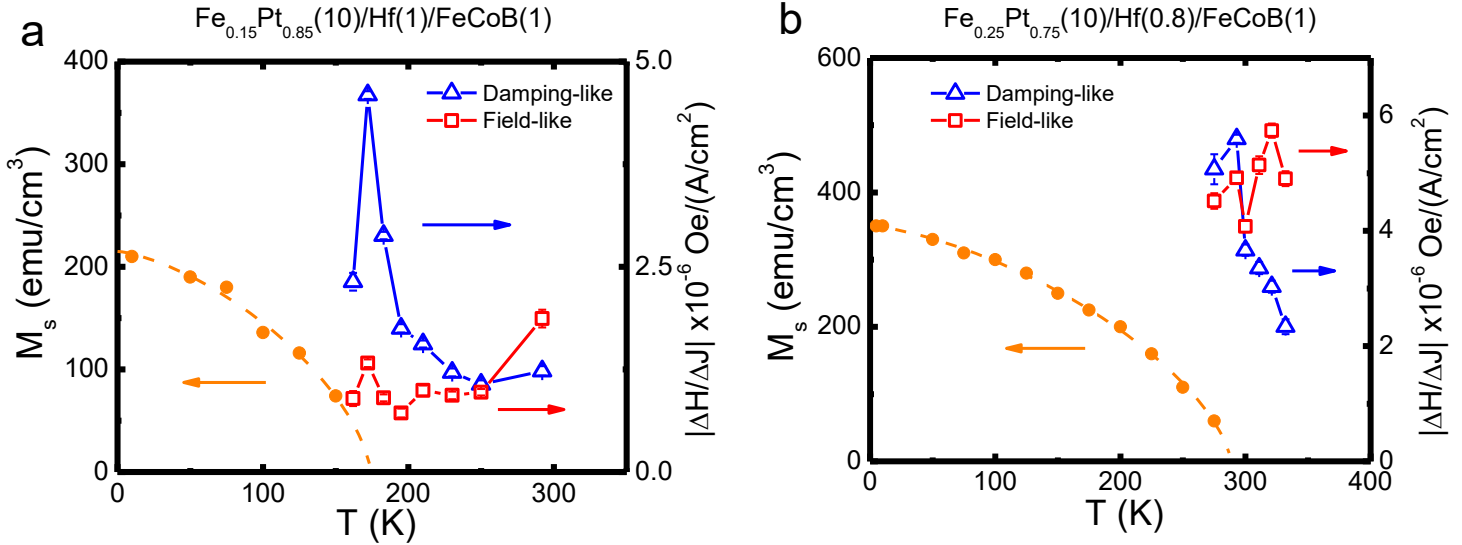

**S2.** Measured values of the damping-like and field-like effective fields as a function of temperature for samples (a) $Fe_{0.15}Pt_{0.85}(10)/Hf(1)/FeCoB(1)$ and (b) $Fe_{0.25}Pt_{0.75}(10)/Hf(0.8)/FeCoB(1)$. Also indicated are the corresponding values of magnetization as a function of temperature for the FePt layers.

In addition to the damping-like effective fields discussed in the main text, we also determined field-like effective fields using the harmonic response technique, for both samples (A) and (B), as shown in Fig. S2a and Fig. S2b. To illustrate the corresponding Curie temperatures, we also plot in both panels the temperature dependence of the magnetization of the appropriate FePt alloy, determined by VSM. In Fig. S2a, for the $Fe_{0.15}Pt_{0.85}$ sample, the field-like effective field decreases as temperature is reduced from 293 K, consistent with the temperature dependence of the field-like term observed previously in pure Pt [1], except that there is a small peak at approximately 170K, coincident with the large peak in the damping-like field. We attribute both peaks to a maximum in the spin current density produced by the FePt alloy. For sample (B) the field-like terms is much stronger overall than that for sample (A), but it also has a local maximum at 293K, coincident with that of the damping-like field. We attribute the more complicated behavior for the field-like effective field evident in Fig. S2b to temperature-dependent changes in the strength of perpendicular anisotropy for the FeCoB layer in sample (B) -- the anisotropy field decreases from 1693 Oe to 909 Oe as T increases from 293K to 330K. As has been previously reported [1] there is in general an apparent inverse correlation between the strength of the field-like term for perpendicularly magnetized HM/FM samples as measured by the harmonic response technique and the FM anisotropy field .

## S3. Second harmonic signals at low temperatures

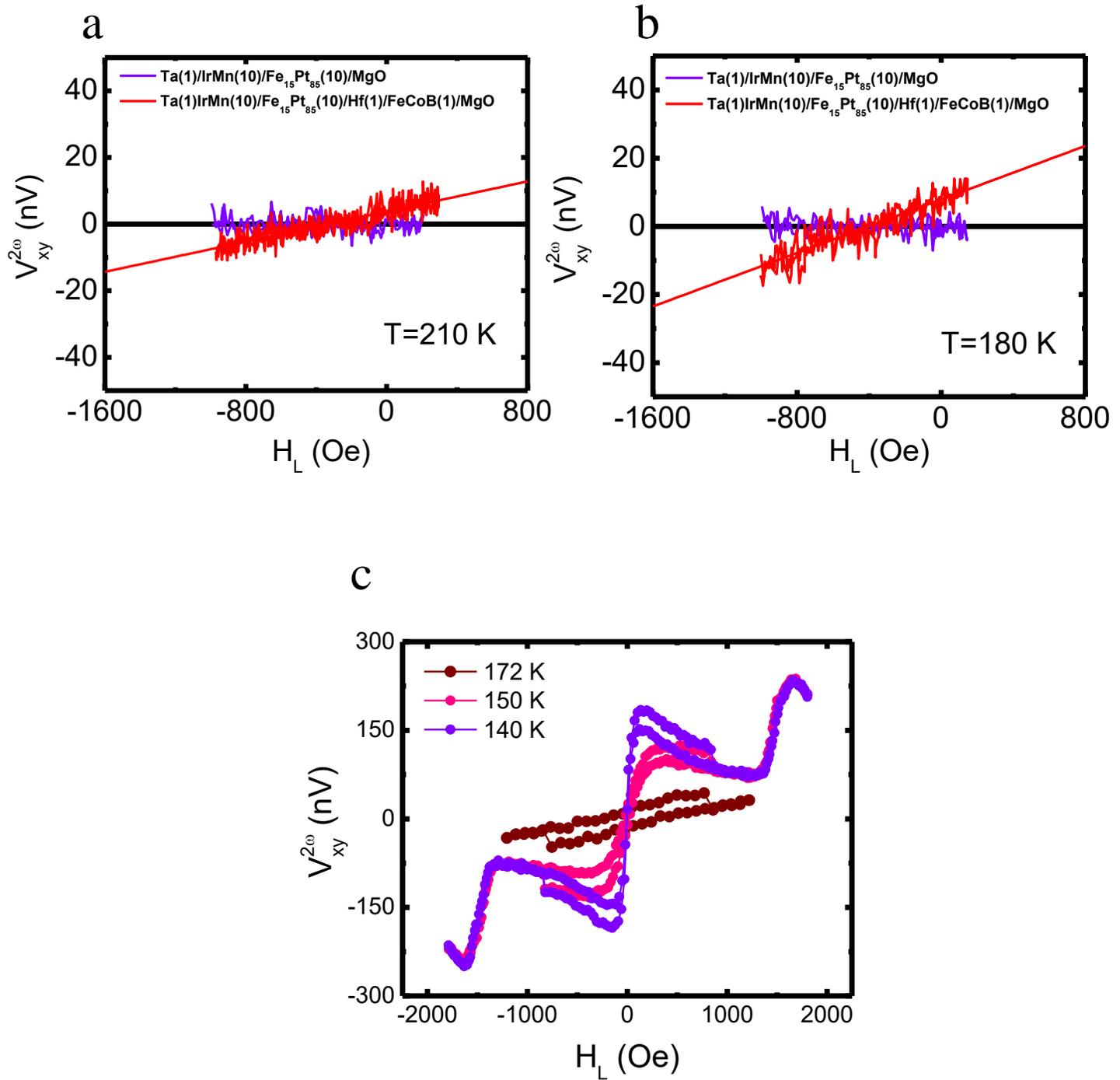

**S3.** (a) and (b) The second harmonic signals of the samples IrMn(10)/Fe$_{15}$Pt$_{85}$(10)/MgO and IrMn(10)/Fe$_{15}$Pt$_{85}$(10)/Hf(1)/FeCoB(1)/MgO in the vicinity of the Curie temperature. The straight lines are linear fit to the data. (c) Second harmonic voltage for sample (A) $Fe_{0.15}Pt_{0.85}$ (10)/Hf(1)/FeCoB(1) below its Curie temperature from 172K to 140K.

In order to confirm that the harmonic measurements in our samples indeed measure the spin torque exerted on the perpendicularly magnetized FeCoB layer due to the spin current from the FePt, instead of a SOT on the FePt exerted by the IrMn layer, we measured the second harmonic signal of a sample without the FeCoB layer, IrMn(10)/$Fe_{15}Pt_{85}$(10)/MgO, in the vicinity of the Curie temperature and compared it with the signal of the sample with the FeCoB layer as in the main text. The results under the longitudinal field sweep are shown in Fig. S3 (a) and (b). It can be clearly seen that in the sample without the FeCoB layer, there is negligible second harmonic signal, while the sample with FeCoB shows an obvious field-dependent signal. This is also true in the transverse field sweep (not shown here). These results demonstrate that the harmonic signals measured in Fig. 2b in the main text are indeed from the perpendicularly magnetized FeCoB layer.

As mentioned in the main text, when the samples are cooled to well below the Curie temperature of the FePt alloys, the emergence of strong ferromagnetism in the FePt layer makes it difficult to measure the spin-orbit effective fields using the harmonic response technique. In Fig. S3 (c) we show as an example the second harmonic voltage for sample (A) measured from 172K down to 140K. For the higher temperature (172K) measurement the data have a simple linear dependence on applied in-plane magnetic field, as expected from the harmonic response technique for a PMA (FeCoB) system in response to spin-orbit torque [2]. The two parallel branches correspond to the up and down magnetization states of the FeCoB layer. This indicates that at 172K, the dominant signal in the second harmonic response is the usual Hall signal from the FeCoB layer. When the heterostructure is cooled to a lower temperature (150K, then to 140 K), there is an additional signal near zero magnetic field that increases in magnitude as temperature is reduced. This extra second harmonic signal, which adds to the FeCoB signal, arises from the now-ferromagnetic $Fe_{0.15}Pt_{0.85}$ layer and is generated on account of the planar Hall effect in the $Fe_{0.15}Pt_{0.85}$ layer as a response to the torques from the other layers. The signature of this effect is a diverging $1/H_L$ field-dependence [2] for the magnitude of the Hall signal near $H_L = 0$, due to deflections of the $Fe_{0.15}Pt_{0.85}$ magnetization within the sample plane, that is distinctly different from the linear field dependence of signals due to the PMA FeCoB layer. One can still clearly see the switching of the FeCoB in the signal at around 700 Oe for all temperatures in Fig. S3 (c), but the strongly nonmonotonic field dependence from the $Fe_{0.15}Pt_{0.85}$ layer obscures the linear signal from the FeCoB layer. As a result, the more-complicated second harmonic signal can no longer be used to make quantitative measurements of spin-orbit torques acting on the PMA FeCoB layer. For the 150K and 140K data at fields above 1000 Oe, the second harmonic signals also begin to deviate from the small angle approximation [3], which is beyond the scope of the work discussed here.

**S4. FMR measurement of FeCoB magnetic damping for in–plane magnetized FePt/Hf/FeCoB multilayers, and determination of the effective spin mixing conductance.**

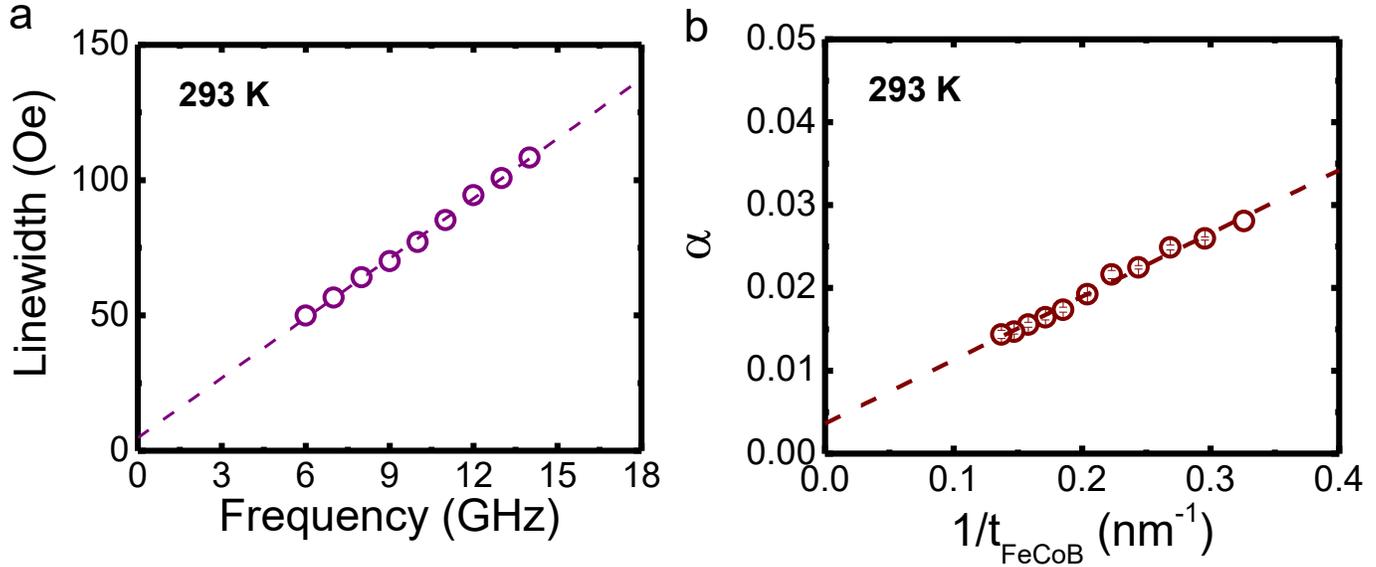

**S4.** (a) Resonance linewidth for sample $Fe_{0.25}Pt_{0.75}$ (10)/Hf(0.25)/FeCoB(7.3) as a function of microwave frequency at 293K. (b) Magnetic damping as a function of the reciprocal of the FeCoB thickness for a series of $Fe_{0.25}Pt_{0.75}$(10)/Hf(0.25)/FeCoB($t_{FeCoB}$) samples.

We used a flip-chip technique to measure the ferromagnetic resonance signals for a series of $Fe_{0.25}Pt_{0.75}$(10)/Hf(0.25)/FeCoB($t_{FeCoB}$)/MgO chips, where $t_{FeCoB}$ ranged from 2.5nm to 7.3nm. In this technique a microwave waveguide optimized for transmission in the 1-20 GHz range carries a 10 dBm rf power generated by a signal generator (Agilent E8257). The sample is placed on top of this waveguide such that the magnetic layers face the waveguide. A dc magnetic field is scanned using an external electromagnet to detect the resonance condition. A small ac field generated by Helmholtz coils, which provides an ac signal for lock-in detection, is added to the dc bias field. When the resonance condition is satisfied, microwave power is absorbed into the uniform precession mode. The changes in the absorbed power (d$P$/d$H$) are detected using a rectifying diode, at the ac field modulation frequency.

Figure S4a shows the resonance linewidth as measured at 293 K as a function of the microwave frequency for the $Fe_{0.25}Pt_{0.75}(10)/Hf(0.25)/FeCoB(7.3)/MgO$ sample, as obtained by fitting the field-swept signal to the derivative of a Lorentzian. The slope of the linear fit to these points gives us the magnetic damping of the free layer. Repeating this measurement for samples with different $t_{FeCoB}$ (Fig. S4b) provides the data needed to obtain $\alpha_0$ and $g_{eff}^{\uparrow\downarrow}$ via the spin pumping theory prediction, $g_{eff}^{\uparrow\downarrow} = 4\pi M_s t_{FeCoB} \Delta\alpha / (\gamma\hbar)$ [4]. The linear fit in Figure S4b gives $\alpha_0 = 0.004 \pm 0.001$ and $g_{eff}^{\uparrow\downarrow} = 65 \pm 2\,nm^{-2}$. The temperature dependence of $g_{eff}^{\uparrow\downarrow}$ was then determined by measuring the enhanced damping via the resonance linewidth of the $Fe_{0.25}Pt_{0.75}(10)/Hf(0.25)/FeCoB(7.3)/MgO$ sample as a function of $T$. Linewidth measurements of a FeCoB layer without an adjacent FePt film showed no significant difference in linewidth over the same temperature range.